\documentclass[useAMS,usenatbib]{mn2e}
\usepackage{graphicx}

\title[On the Erigone family and the $z_2$ secular resonance]
{On the Erigone family and the $z_2$ secular resonance}
\author[V. Carruba, S. Aljbaae, and O. C. Winter]{V. 
Carruba$^{1,2}$\thanks{E-mail: vcarruba@feg.unesp.br}, S. Aljbaae$^{1}$, 
and O. C. Winter$^{1}$\\
$^{1}$UNESP, Univ. Estadual Paulista, Grupo de din\^{a}mica Orbital e
  Planetologia, Guaratinguet\'{a}, SP, 12516-410, Brazil. \\
$^{2}$Department of Space Studies, Southwest Research Institute, Boulder, 
CO, 80302, USA.\\
}

\begin{document}

\date{Accepted 2015 October 19.  Received 2015 October 19; in original form 2015 August 03.}

\pagerange{\pageref{firstpage}--\pageref{lastpage}} \pubyear{2015}

\maketitle

\label{firstpage}

\begin{abstract}
The Erigone family is a C-type group in the inner main belt.  Its
age has been estimated by several researchers to be less then 300 My,
so it is a relatively young cluster.  Yarko-YORP Monte Carlo methods 
to study the chronology of the Erigone family confirm results obtained 
by other groups.  The Erigone family, however, is also characterized by its 
interaction with the $z_2$ secular resonance.  While less than 15\% of 
its members are currently in librating states of this resonance, the number 
of objects, members of the dynamical group, in resonant states is 
high enough to allow to use the study of dynamics inside the $z_2$ 
resonance to set constraints on the family age.  

Like the ${\nu}_{6}$ and $z_1$ secular resonances, the $z_2$ resonance is 
characterized by one stable equilibrium point at $\sigma = 180^{\circ}$ in 
the $z_2$ resonance plane $(\sigma, \frac{d\sigma}{dt})$, where $\sigma$ is the
resonant angle of the $z_2$ resonance.  Diffusion in this plane occurs on 
timescales of $\simeq 12$ My, which sets a lower limit on the Erigone family 
age. Finally, the minimum time needed to reach a steady-state population
of $z_2$ librators is about 90 My, which allows to impose another, 
independent constraint on the group age.  
\end{abstract}

%

\section{Introduction}
\label{sec: intro}

The Erigone family is a C-type group in the inner main belt. Its 
age has been estimated to be $280^{+30}_{-50}$~My
by Vokrouhlick\'{y} et al. (2006c) with a Yarkovsky and 
Yarkovsky -O'Keefe -Radzievsky -Paddack (YORP, overall we defined this
method as Yarko-Yorp) Monte Carlo
approach, $170^{+25}_{-30}$~My by Bottke et al. (2015) with a modified 
Yarko-Yorp Monte Carlo approach that also includes the effect of stochastic
YORP, and as  $220^{+60}_{-80}$ My in Spoto et al. (2015).  There seems
therefore to be a significant consensus in the literature that the Erigone 
family should be a relatively young C-type cratering family.
Much less attention has been given to its secular properties.
This group is characterized by its interaction with the $z_2$ secular 
resonance (Carruba and Michtchenko 2009), a resonance of the $(2g+s)$-type
introduced by Milani and Kne\v{z}evi\'{c} (1994) using the perturbation
theory of Milani and Kne\v{z}evi\'{c} (1990)\footnote{Secular resonances 
involve commensurabilities between the asteroid proper frequency of 
precession of argument of pericenter $g$ and longitude of the node $s$ 
and that of planets.  The $z_2$ secular resonance is a non-linear secular 
resonance that involves the combination of the two linear resonances ${\nu}_6 =
g-g_6$ and ${\nu}_{16} = s-s_6$, where the subscript 6 is associated 
with the sixth planet, Saturn.  In particular, $z_2 = 2{\nu}_6+{\nu}_{16}=
2(g-g_6)+(s-s_6)$.}.  About 15\% of the current Erigone family members
are in librating states of the $z_2$ secular resonance.  While this is
not a majority of the members, as for the case of the Tina family and
the ${\nu}_6$ secular resonance (100\%, Carruba and Morbidelli 2011),
the Agnia and Padua families, with more than 50\% of the members in 
$z_1$ librating states (Vokrouhlick\'{y} et al. 2006b, Carruba 2009), 
dynamics inside the $z_2$ secular resonance can still be used 
to obtain estimates, or at least lower limits, of the family age.

Also, dynamics inside the $z_2$ secular resonance has not so far been given
a lot of attention.  As for the ${\nu}_6$ and $z_1$ secular resonances, 
we found an equilibrium point in the $(\sigma, \frac{d\sigma}{dt})$ plane
at $\sigma = 180^{\circ}$, where $\sigma$ is the resonant argument of the 
$z_2$ secular resonance.  The minimum time needed to fully disperse an
originally compact family in this plane fully around the equilibrium
point can be used to set constraint on the group age.  Furthermore, checking 
the minimum time needed to inject a steady-state librating population 
into the $z_2$ secular resonance can also provide constraints on the Erigone 
family age of a nature not previously used in the literature.

This paper is so divided: in Sect.~\ref{sec: prop_el} we obtain dynamical 
maps and synthetic proper elements for asteroids in the orbital region of
Erigone.  Sect.~\ref{sec: sec_dyn} deals with secular dynamics in the area. 
In this section we estimate the population of likely resonators of the
main secular resonances in the region (Carruba 2009), and verified
their resonant behavior by studying their resonant arguments.  
Sect.~\ref{sec: taxonomy} revisites physical properties of the local
asteroids such as taxonomy, geometric albedo, and estimated masses.  In
Sect.~\ref{sec: fam_ide} we identify the Erigone family in various domains
of proper elements, frequency, and photometric colors.  In 
Sect.~\ref{sec: chron} we use Yarko-Yorp Monte Carlo method to obtain a 
preliminary estimate of the family age.  In Sect.~\ref{sec: z2_dyn}
we investigate the dynamics inside the $z_2$ secular resonance and
the constraints that it may provide on the age of the Erigone family.
Sect.~\ref{sec: dyn_evol} deals with the dynamical evolution of the Erigone 
family, simulated with the SYSYCE integrator of Carruba et al. (2015).
Minimum times needed to inject into the $z_2$ secular resonance a 
steady-state population allowed to set further
constraints on the Erigone family age.  Finally, in Sect.~\ref{sec: conc} 
we present our conclusions.

\section{Proper elements}
\label{sec: prop_el}

To study the local dynamics we first obtained a proper elements 
map in the region of the Erigone family.  We computed synthetic proper 
elements with the method discussed in Carruba (2010) for a grid
of 51 by 75 (i.e, 3825) test particles in the $(a,sin(i))$ plane.
Values of proper $i$ ranged from $3.80^{\circ}$ to $6.76^{\circ}$, while 
values of proper $a$ were in the range from 2.3 to 2.5 AU. 
All the other initial angles and the eccentricity were 
those of (4) Erigone at J2000.

\begin{figure}
  \centering
  \centering \includegraphics [width=0.45\textwidth]{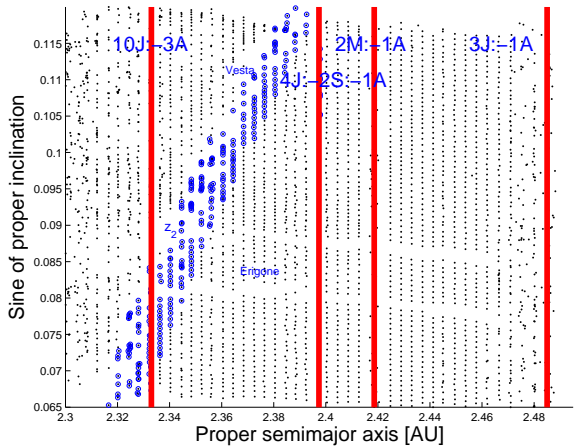}

\caption{Dynamical map in the proper $(a,sin(i))$ space for the orbital
region of the Erigone family.  Vertical red lines display the positions of the 
main mean-motion resonances, blue circles identify $z_2$ likely
resonators.}
\label{fig: map_erigone}
\end{figure}

Results are shown in Fig.~\ref{fig: map_erigone}, where each black dot
identifies the proper elements of a given test particle.  Mean-motion
resonances are shown as vertical red lines in the figure, while 
secular resonances appear as inclined bands of lower number density of proper 
elements.  We identify all mean-motion
resonances up to order 13 and all linear secular resonances (and some 
non-linear) in the system.   The Erigone family is characterized by its
interaction with the $z_2 = 2(g-g_6)+s-s_6$ secular resonance, 
that appears in the map as an inclined band at approximately 
$a = 2.35$~AU.  We also plotted as blue circles particles
whose value of $2g+s$ is in the range of $2g_6+s_6 \pm 0.3$ arcsec/yr,
and that are likely to be found in $z_2$ librating states (see 
Sect.~\ref{sec: sec_dyn} for further details).
Other non-linear secular resonances in the region   
are the $g-$type $3{\nu}_6-{\nu}_5$ resonance and the $g+2s$-type
${\nu}_5+2{\nu}_{16}$, plus harmonics involving combination of other
planets proper frequencies as Uranus.

We then computed proper elements for numbered objects in the region of
the Erigone family, defined as a box in the $(a,e,sin(i))$ space whose 
limits are given by the maximum and minimum proper elements of the Erigone
family halo, as obtained in Carruba et al. (2013).  Of the 4956 asteroids
that we integrated,  whose osculating elements were downloaded from the 
AstDyS site (http://hamilton.dm.unipi.it/astdys, Kne\v{z}evi\'{c} and 
Milani 2003) on June 24th 2014, all survived a 10 My integration.
We computed synthetic proper elements,
and we eliminated all objects for which one of the proper elements 
$a, e, \sin{(i)}$ or proper frequencies $g$ and $s$ had errors larger than 
those classified as ``pathological'' by Kne\v{z}evi\'c and Milani (2003); i.e., 
$\Delta a = 0.01$~AU, $\Delta e = 0.1$, $\Delta \sin{(i)} = 0.03$, 
and $\Delta g = \Delta s = 10~arcsec/yr$.   This left us
with a data-set of 4717 asteroids with good-quality proper elements.
 
The Erigone family appears as a two-lobed structure among asteroids with stable
elements. The main sources of instabilities on proper $a$ in the Erigone 
region are associated with the 10J:-3A, 4J:-2S:-1A and 2M:-1A 
mean-motion resonances.  Other minor three-body
and four-body resonances are not shown in Fig.~\ref{fig: map_erigone}
for simplicity.  Secular dynamics in the Erigone area will be discussed 
in the next section.

\section{Secular dynamics in the Erigone region}
\label{sec: sec_dyn}

Secular resonances occur when there is a commensurability between 
the proper frequency of precession of the argument of pericenter $g$
or of the longitude of the node $s$ of a given asteroid and a planet.
If the commensurability is just between one frequency of the asteroid
and one frequency of the planet, such as in the case of 
the ${\nu}_6 = g-g_6$ secular resonance, the resonance is linear.
Higher order resonances involving more complex combinations 
of asteroidal and planetary frequencies, such as for instance
the $z_1= {\nu}_6 +{\nu}_{16} = g-g_6+s-s_6$ resonance, are called non-linear 
secular resonances.  Carruba (2009) defines as likely resonators
the asteroids whose frequency combination is to within $\pm 0.3$ arcsec/yr
from the resonance center for the case of the $z_1$ secular resonance.  
For the the $z_2$ resonance,
this means that $2g+s = 2g_6+s_6 = 30.345$~arcsec/yr (we are using
values of the planetary frequencies from Milani and Kne\v{z}evi\'{c} 1994,
see also table 1 in Carruba and Michtchenko 2007).  About 90\% of the
$z_1$ likely resonators in the Padua family area were found to be
actual librators, when the resonant argument of the resonance was analyzed
(librators were defined as objects whose resonance argument was
observed to librate for the whole length of the numerical simulation
performed to check their status, i.e., 20 Myr)\footnote{Since the 
resonances studied in this work are of higher order and degree then 
the $z_1$, their actual width should be smaller.  Using the 0.3 arcsec/yr 
criteria could therefore overestimate a bit the number of actual 
resonators.  Yet, since the actual width of a given resonance is 
difficult to predict before an analysis in $(\sigma, \frac{d\sigma}{dt})$ 
plane of each resonance is performed, we believe that using the 0.3 arcsec/yr 
should be a reasonable first-order approximation.}.

\begin{table*}
\begin{center}
\caption{{\bf Main secular resonances in the Erigone region labelled 
in terms of combination of linear resonances and proper frequencies, 
frequency value, number of likely and actually resonant asteroids.}}
\label{table: sec_res}
\vspace{0.5cm}
\begin{tabular}{|c|c|c|c|c|}
\hline
                   &                    &           &            &           \\
Resonance argument & Resonance argument & Frequency & Likely     & Actual    \\
in terms of        & in terms of        & value     & resonators & resonators\\
linear resonances  & proper frequencies & $[``/yr]$ &            &           \\
                   &                    &           &            &           \\
\hline
                   &                    &           &            &           \\
g resonances       &                    &           &            &           \\
$3{\nu}_6-{\nu}_5$ &  $2g-3g_6+g_5$      & 40.236    &     397    &     75    \\
g+2s resonances    &                    &           &            &           \\
${\nu}_5+2{\nu}_{16}$& $g+2s-g_5-2s_6$    &-48.433    &     192   &      38    \\
2g+s resonances    &                    &           &           &            \\
$z_2=2{\nu}_6+{\nu}_{16}$& $2g+s-2g_6-s_6$& 30.345    &     493   &     367    \\
                   &                    &           &            &           \\
\hline
\end{tabular}
\end{center}
\end{table*}

Table~\ref{table: sec_res} displays the main secular resonances involving
Jupiter and Saturn frequencies in the region labelled 
in terms of combination of linear resonances and proper frequencies, 
the values of the asteroid resonant frequencies, the number of likely 
resonators, and the number of objects for which the resonant argument 
was actually in libration.  Local secular resonances involving Martian 
frequencies have been discussed in Carruba et al. (2005).  Interested 
readers could find more information in that paper.   For simplicity, we 
do not show in the table resonances involving Uranus frequencies.  
Since the $g_7$ frequency is quite close to the $g_5$, resonances 
involving $g_7$ are very close in proper element space to those 
involving $g_5$, and of lower strength. 

\begin{figure*}

  \centering
  \begin{minipage}[c]{0.5\textwidth}
    \centering \includegraphics[width=3.5in]{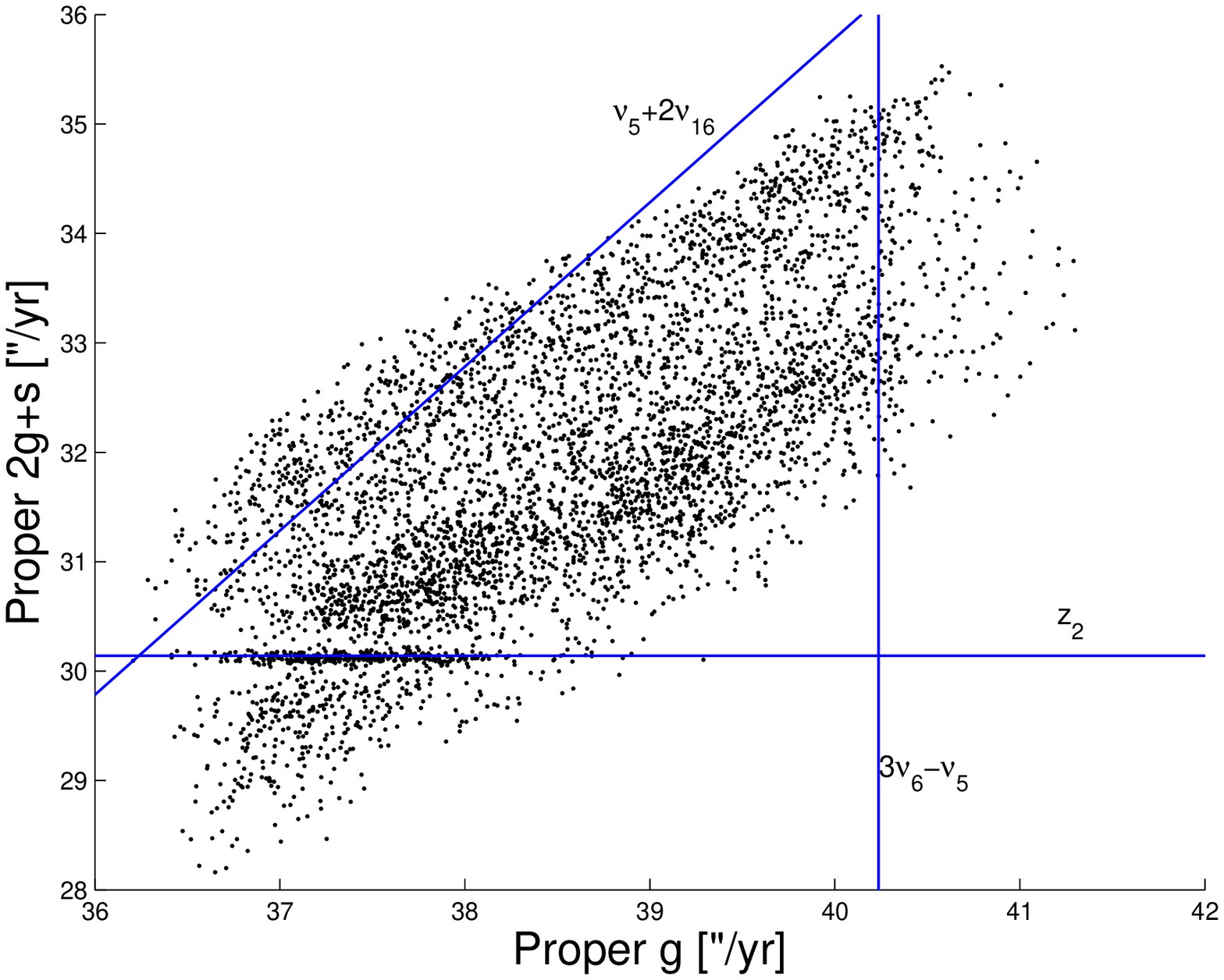}
  \end{minipage}%
  \begin{minipage}[c]{0.5\textwidth}
    \centering \includegraphics[width=3.5in]{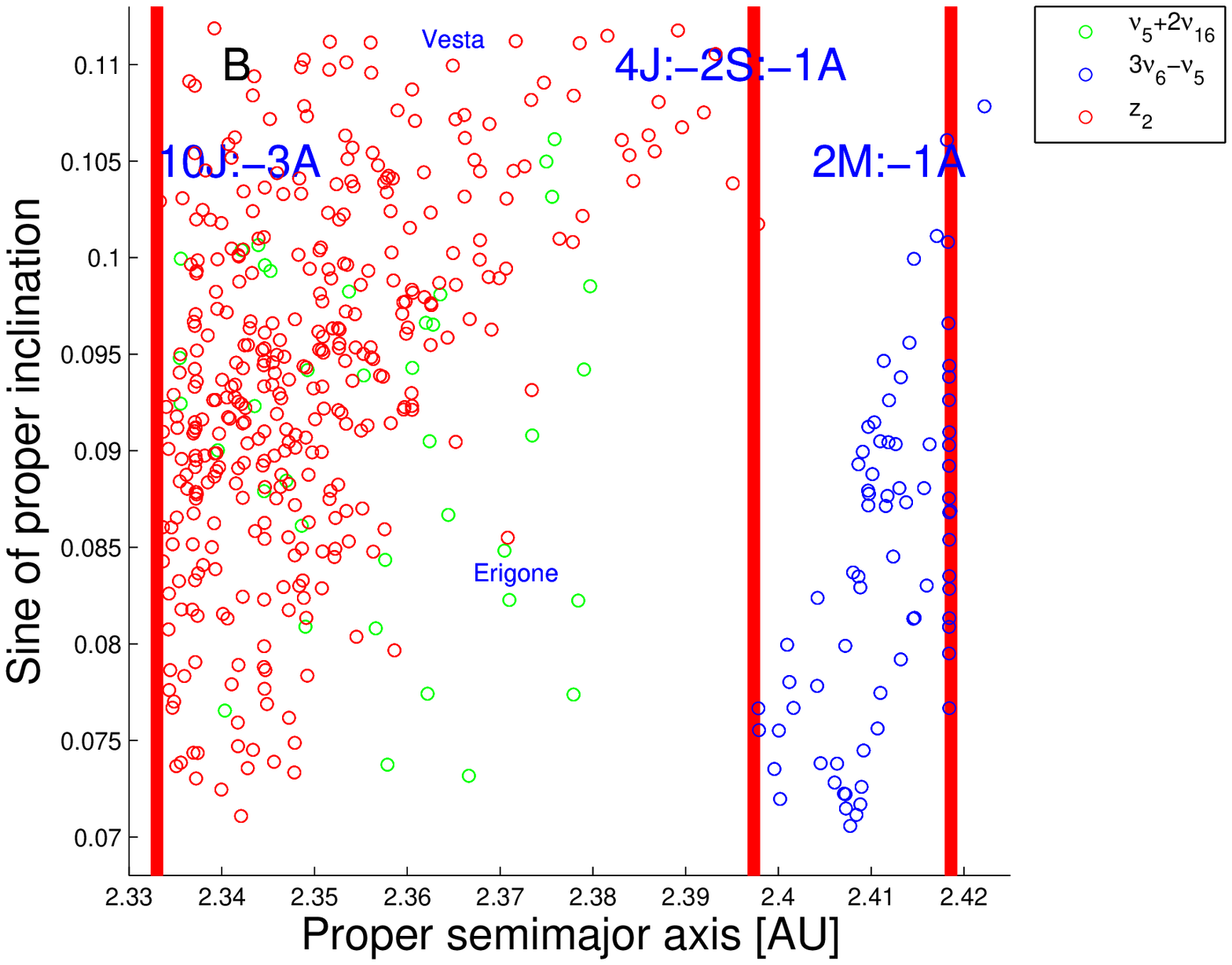}
  \end{minipage}

\caption{Panel A: a (g,2g+s) projection of the 4171 asteroids with known proper
elements in our sample.  Blue lines identify the location of the secular
resonances listed in Table~\ref{table: sec_res} with a number
of ``likely resonators'' larger than 1.  Panel B: an (a,sin(i)) projection
of the actual resonators population listed in Table~\ref{table: sec_res}.} 
\label{fig: erigone_ggs}
\end{figure*}

Fig.~\ref{fig: erigone_ggs}, panel A, displays a (g,2g+s) projection of the 
4171 asteroids with known proper elements in our sample.  Blue lines 
identify the location of the secular resonances listed in 
Table~\ref{table: sec_res}.  Note the horizontal alignment of objects 
along the $z_2$ secular resonance, suggesting the dynamical importance of this 
commensurability.   Panel B displays an (a,sin(i)) projection
of the actual resonators population.  Some of the asteroids in 
$3{\nu}_6-{\nu}_5$ librating states are also inside
the 2M:-1A mean-motion resonance, whose dynamics has been studied
in Gallardo et al. (2011).   Despite all three studied
resonances being of order six, the $z_2$ resonance plays by far
the major role in dispersing asteroids in the Erigone area.  
This may be caused by the fact that the torsions of the secular
resonances, even if associated with terms of the same order, can be 
different by an order of magnitude (Milani and Kne\v{z}evi\'{c}
1994).  The role of the $z_2$ resonance in affecting the evolution of 
members of the Erigone family will
be investigated in further detail in later sections.  Here we start
by having a closer look at the main frequencies involved in this dynamics.

\begin{figure*}

  \centering
  \begin{minipage}[c]{0.5\textwidth}
    \centering \includegraphics[width=2.5in]{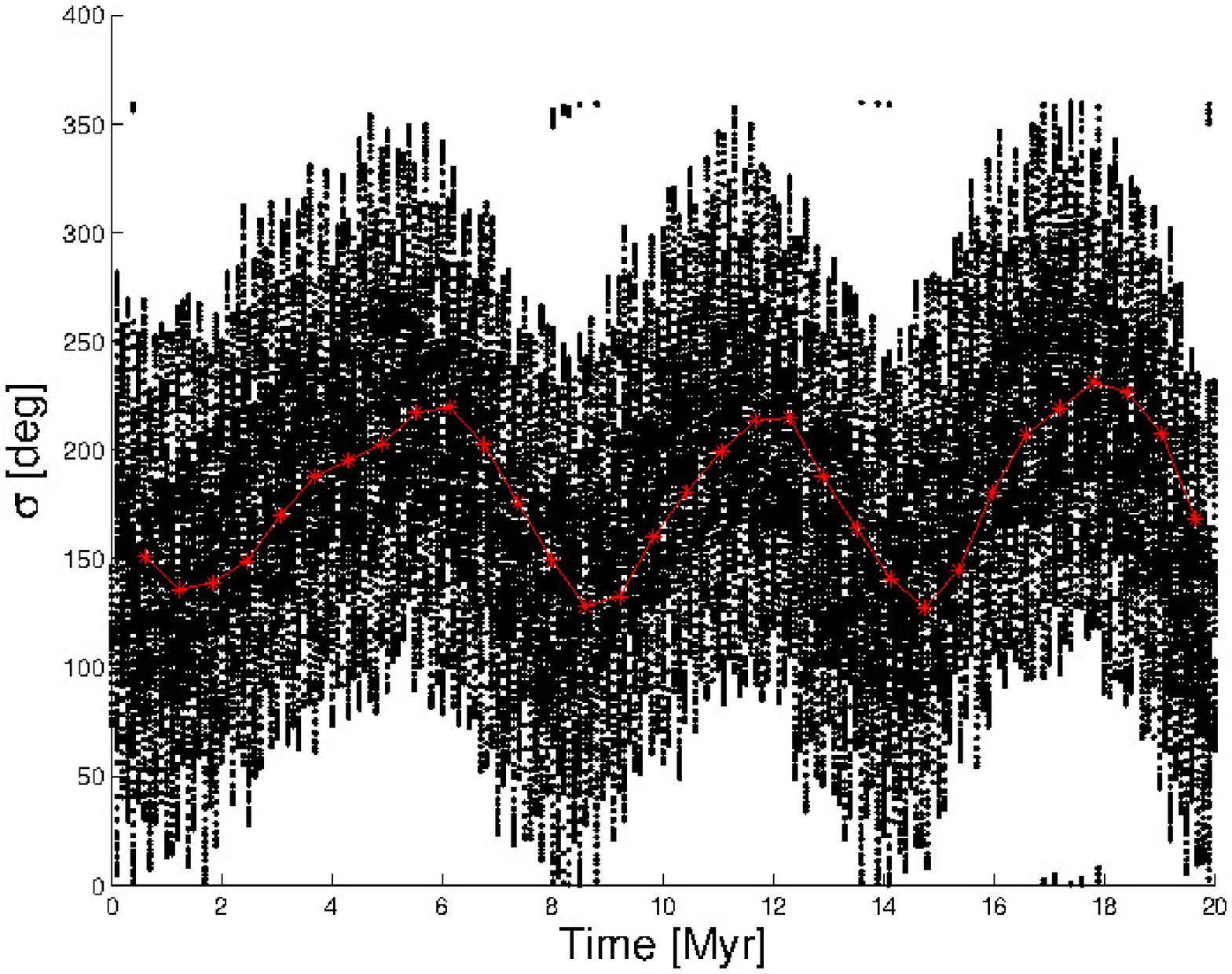}
  \end{minipage}%
  \begin{minipage}[c]{0.5\textwidth}
    \centering \includegraphics[width=2.5in]{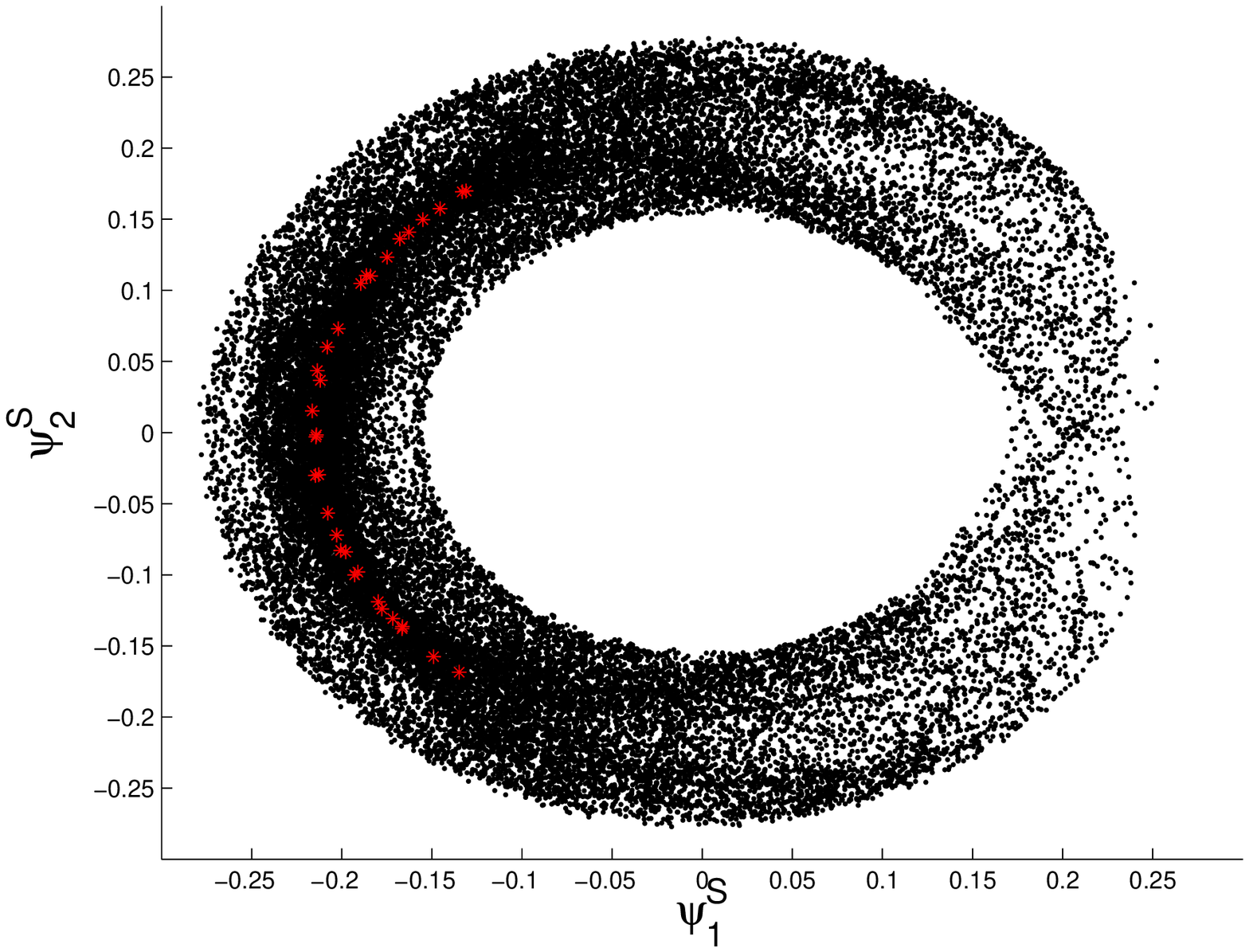}
  \end{minipage}

\caption{Panel A: the time evolution of the resonance argument 
$\sigma=2({\varpi}-{\varpi}_6)+(\Omega-{\Omega}_6)$ (black dots).  
The red asterisks show the argument after 
being digitally filtered so as to remove all frequencies corresponding to
periods less than 700,000 yr.  Panel B:  Orbital evolution in the 
polar plane $({{\psi}_1}^S, {{\psi}_2}^S)$. The resonant 
elements $({{\psi}_1}^S, {{\psi}_2}^S)$ were also digitally 
filtered with the same procedure (red asterisks).}
\label{fig: erigone_arg}
\end{figure*}

Fig.~\ref{fig: erigone_arg}, panel A, shows the time behavior of the 
resonant argument $\sigma =2({\varpi}-{\varpi}_6)+(\Omega-{\Omega}_6)$ 
for (8089) Yukar, the lowest numbered among the $z_2$ resonant asteroids 
(black dots).  To emphasize the long-time behaviour of the angle and 
eliminate all short period perturbations such as the $g_5$ frequency in 
the precession of Saturn's pericenter, the argument $\sigma$ was digitally 
filtered, so to eliminate all frequencies with period less than 700,000 
yr (Carruba et al. 2005). Results are shown as red asterisks.  

The reader may notice that there are two main frequencies, one associated
with short-period oscillation around the center of libration, with a 
period of $\simeq 0.06$ My, and another with oscillations of the center 
of libration itself, with a period of $\simeq 6.1$ My.  
Fig.~\ref{fig: erigone_arg}, panel B, shows the 
orbital evolution in a polar diagram, with axes 
$({{\psi}_1}^S, {{\psi}_2}^S)$ defined as:

\begin{equation}
{{\psi}_1}^S = \sqrt{2 (1-\sqrt{1-e^2})} \cos{[2({\varpi}-{\varpi}_6)+(
\Omega -{\Omega}_6)]},\\
\label{eq: psi_1}
\end{equation}

\noindent and, 

\begin{equation}
{{\psi}_2}^S = \sqrt{2 (1-\sqrt{1-e^2})} \sin{[2({\varpi}-{\varpi}_6)+(
\Omega -{\Omega}_6)]},\\
\end{equation}

\noindent where $e$ is the asteroid eccentricity.  Charlier (1902)
introduced these variables (which are asymptotically equal to
$e\cos\sigma$ and $e\sin\sigma$ for small $e$) in his theory for
secular resonances.  Again, black dots show the evolution with orbital
elements filtered up to a period of 700 yr, while the curve shows the
evolution with elements filtered up to a period of 700,000 yr.  The
red asterisks describe an arc associated with a banana-shape orbit, 
visible at higher resolutions than that used for Fig.~\ref{fig: erigone_arg}, 
panel B, with period of about 6.1 My, a similar behaviour is shared by 
other resonant asteroids in the region.
 
Having examined the effect of local dynamics, we are now ready to 
analysize the taxonomical properties of asteroids in the Erigone region.

\section{Compositional analysis: Taxonomy and physical properties}
\label{sec: taxonomy}

Only five objects had taxonomical data in three major photometric/spectroscopic
surveys (ECAS (Eight-Color Asteroid Analysis, Zellner et al. 1985; 
Tholen 1989), SMASS (Small Main Belt Spectroscopic Survey, 
Xu {\em et al.} 1995, Bus and Binzel 2002a,b), and S3OS2 (Small Solar 
System Objects Spectroscopic Survey, Lazzaro et al. 2004):
60 Echo (S-type), 163 Erigone (C), 571 Dulcinea(S), 2763 Jeans (V), 
and 2991 Bilbo (C). Using the classification method of De Meo and Carry 
(2013) that employs Sloan Digital Sky Survey-Moving Object Catalog data, 
fourth release (SDSS-MOC4 hereafter, Ivezic et al. 2001) to compute $gri$ 
slope and $z' -i'$ colors, we obtained a set of 470 observations of asteroids 
(including multiple observations of the same object) in the Erigone region.
This corresponds to 276 asteroids for which a SDSS-MOC4 
taxonomical classification and proper elements are both available. We 
found 30 X, 7 D, 103 C, 52 L, 66 S, 3 V, and 5 B type objects, respectively.
There were data for 18 objects currently inside the $z_2$ secular resonance,
6 were L-, 4 S-, 4 C-, 2 K-, and 2 B-types. 

\begin{figure*}

  \centering
  \begin{minipage}[c]{0.5\textwidth}
    \centering \includegraphics[width=3.5in]{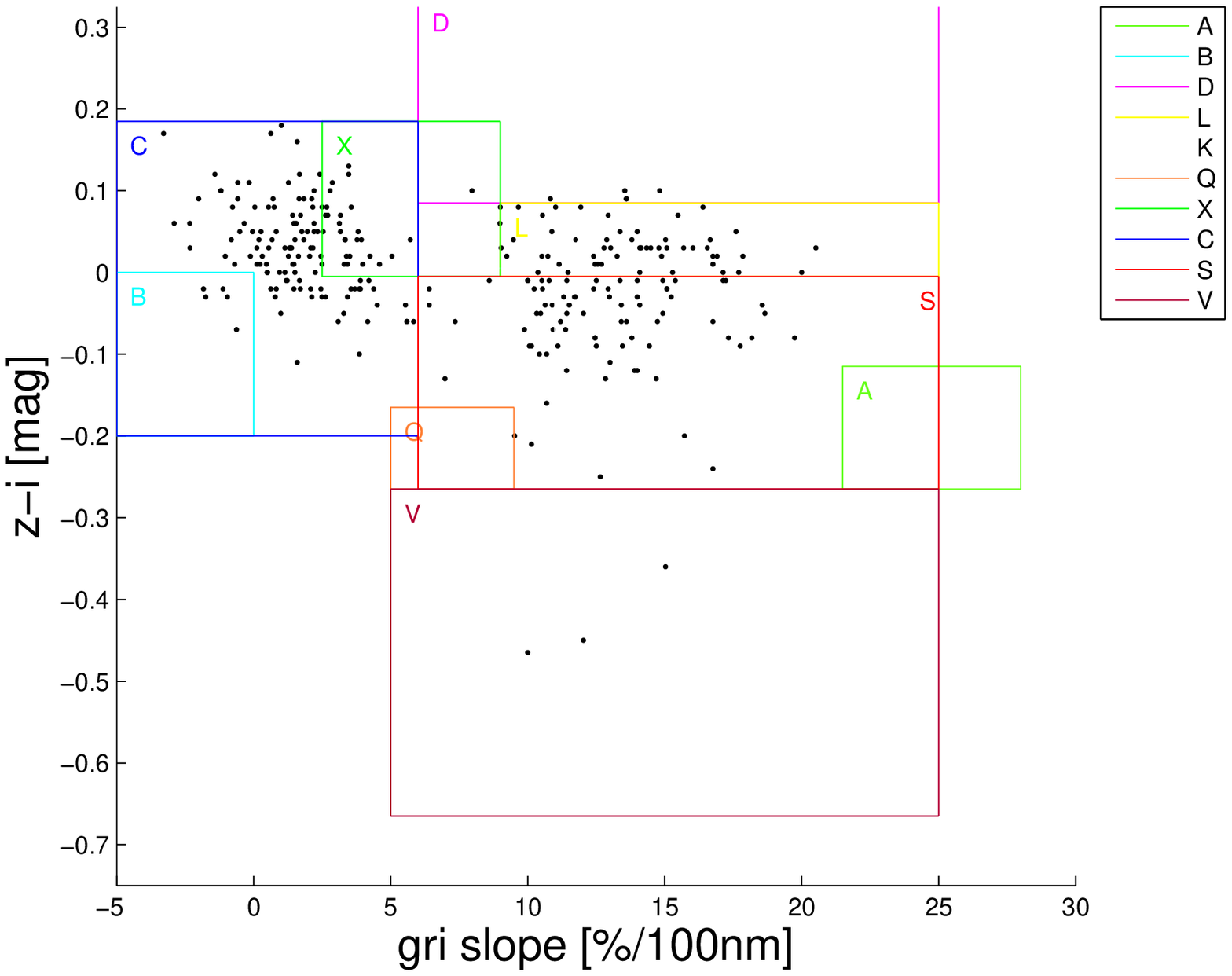}
  \end{minipage}%
  \begin{minipage}[c]{0.5\textwidth}
    \centering \includegraphics[width=3.5in]{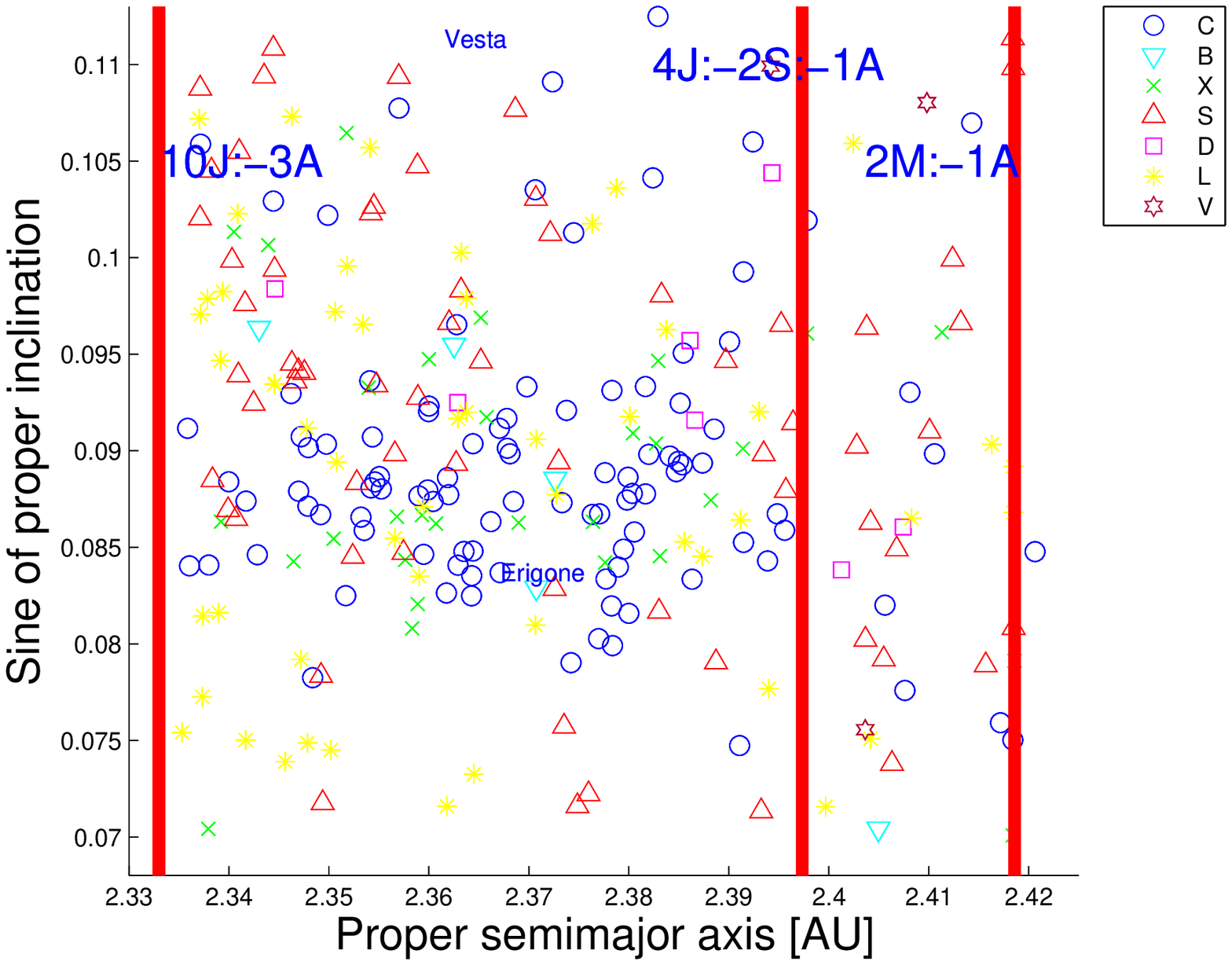}
  \end{minipage}

\caption{Panel A: a projection in the $gri$ slope versus $z' -i'$ plane of
the 276 observations in the SDSS-MOC4 catalogue for asteroids in the
Erigone region.  Panel B: an $(a,sin(i)$ projection of the 280 asteroids
with taxonomical information in the same area.} 
\label{fig: erigone_sdss}
\end{figure*}

Fig.~\ref{fig: erigone_sdss}, panel A, displays a projection in the $gri$ slope 
versus $z' -i'$ plane for the 276 observations in the SDSS-MOC4 catalogue 
for asteroids in the Erigone region, while panel B shows an $(a,sin(i))$ 
projection of asteroids in the same region.  As in Carruba et al. (2013),
that, however, used a simplified method allowing only to distinguish if 
an asteroid belonged to the C-complex, S-complex, or was a V-type,
we found that the Erigone region is dominated by the C-type Erigone family,
but with significant mixing of other S and V type objects.

Concerning albedo information, we identified 1053 asteroids with 
WISE  albedo information (Masiero et al. 2012) in the Erigone region. 
Our analysis confirms that of Carruba et al. (2013): the Erigone region
has a predominance of lower albedos, with 802 asteroids (76.2\% of the total) 
with $p_V < 0.15$, associated with C-complex objects, and 251 
(23.84\% of the total) bodies with $p_V > 0.15$, associated with S-complex and 
V-type objects\footnote{The boundary between C- and S-complex classes
in albedo varies in the literature.  While most of C-type objects have
values of albedos less than 0.1, recent results from WISE showed
that some C-type families, such as 2782 Leonidas, 1128 Astrid, 
and 18405 FY12, have significant tails at higher albedos than previously 
thought.  Also, WISE albedo data has usually errors of the order of 
10\%, 20\% for  $p_V \simeq 0.1$.   In this work we prefer to exclude 
less objects by using a higher threshold ($p_V =0.15$), rather than run 
the risk of eliminating possible family members.}.  Using the values of 
the diameters from WISE, when available (otherwise diameters are estimated 
using absolute magnitudes and the mean value of geometric albedo in the 
Erigone region, i.e., $p_V =0.033$ via Eq.~4 in Carruba et al. 2003), and 
the density of 163 Erigone from Carry (2012), we computed the masses of 
asteroids in the Erigone region, assumed as homogeneous spheres.  For the 
few asteroids where an estimate of the mass was reported in Carry (2012), we 
used the values from that paper.   The only two objects with a mass larger 
than $10^{17}$~kg in the region are 163 Erigone itself, and the family-less 
S-type 60 Echo.  This would suggest that the only large C-type family in the
region should be the one associated with 163 Erigone.  We will 
further investigate this issue in the next section.

\section{Family identification}
\label{sec: fam_ide}

To obtain family membership of the Erigone family 
(FIN 406, according to Nesvorn\'{y} et al. 2015), we first followed
the approach of Carruba et al. (2015), where the hierarchical clustering
method was used in a domain of proper elements $(a,e,sin(i))$ and 
$gri$ slope and $z' -i'$ colors from SDSS-MOC4, in the restricted 
sample (280 asteroids, 5.94\% of the 4717 asteroids in the area), 
for which this information is available.  While the sample of asteroids 
in this domain is reduced, this method is more efficient in eliminating 
possible interlopers then standard HCM in the domain of proper 
(a,e,sin(i)) only, and can provide clues about the possible existence 
(or not) of large families in the area.  The value of the minimal distance 
cutoff $d_0$ defined as in Beaug\'{e} and Roig (2001)was of 154.5~m/s and 
the minimal number of objects to have a group statistically significant 
was 25.   In this five-dimension domain the Erigone family merged
with the local background for a cutoff of 225 ~m/s, and we did not find
any other significant groups in the region, so confirming the 
hints provided by the analysis of asteroid masses in the region in 
the previous section.  Following Spoto et al. (2015), we considered 
the groups around 163 Erigone and 5026 Martes as belonging to the 
same dynamical family.  

\begin{figure}

\centering
\centering \includegraphics [width=0.45\textwidth]{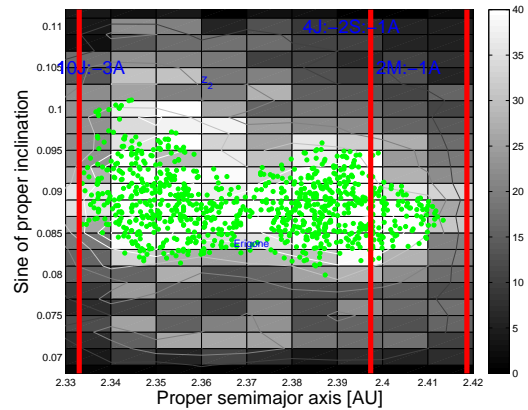}

\caption{Contour plot of number density of asteroids
in the same region.  Green dots show the location of members
of the identified Erigone family.}
\label{fig: erigone_dens}
\end{figure}

We then used a standard HCM approach in the $(a,e,sin(i))$ domain,
with the methods described in Bendjoya and Zappal\`{a} (2002) for the
4717 asteroids with synthetic proper elements in the area.  As in Carruba et 
al. (2013), we decided to work with a family obtained at a cutoff of
40~m/s.  After eliminating 9 asteroids with incompatible taxonomy, and 
8 objects with albedo $p_V > 0.15$, we were left with a family of 880
members.  Fig.~\ref{fig: erigone_dens} displays a contour plot of 
number density of asteroids in the Erigone region, where green full dots 
identify the orbital location of the family members identified in this work.
We plotted a contour plot of the number density of background asteroids,
rather than showing the location of each asteroid orbit as a dot, so
as to facilitate the visual identification of the Erigone family members.
We computed the number of asteroids per unit square in a 30 by 30 grid in 
the $(a,sin(i))$ plane, with $a$ between 2.3 and 2.45 and $sin(i)$ 
between 0.065 and 0.95.  Regions with lower number of 
asteroids are usually (but not always) associated with dynamically unstable
region.  Number density of asteroids may therefore provide clues
about the local dynamics.  As discussed before, no other significant group
was identified in the Erigone area.

We identified 123 members of the Erigone dynamical family
in librating states of the $z_2$ secular resonance, but only five
members in $3{\nu}_6-{\nu}_5$ librating states, and none in 
${\nu}_5+2{\nu}_{16}$ librating states.  Because of the very low numbers
of resonant objects (less then  0.6\% of the $(a,e,sin(i))$ family at most), no 
statistically significant information can be unfortunately derived from 
an analysis of the dynamical evolution of the resonant populations of 
the latter two resonances.

\begin{figure}
\centering
\centering \includegraphics [width=0.45\textwidth]{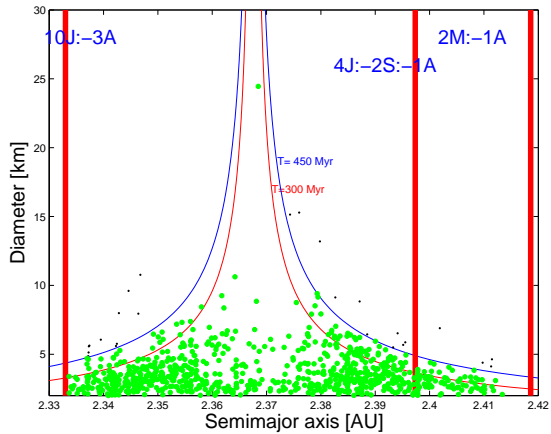}

\caption{A proper $a$ versus diameter projection of 
members of the Erigone family.  The red and blue curves
display isolines of maximum displacement in $a$ caused by Yarkovsky
effect after 300~My (red line) and 450~My (blue line).  Green full dots 
are associated with Erigone family members, black dots display the 
location of dynamical interlopers.}
\label{fig: iso_yarko}
\end{figure}

As a next step in our research, we then obtained a preliminary estimate
of the age of the families and attempted to eliminate possible 
dynamical interlopers using the method of Yarkovsky isolines.  Following the 
approach of Carruba et al. (2015), we computed isolines of displacements caused
by the Yarkovsky effect for a C-type family.  Dynamical interlopers 
are objects that reside beyond the maximum possible Yarkovsky isoline, and 
could not have reached their current orbital position since the family 
formation.  Fig.~\ref{fig: iso_yarko} displays the results of our method.
Overall, we found 26 asteroids that could be classified as dynamical
interlopers, leaving us with a family of 854 objects.

In the next section we will use the so-called Yarko-YORP Monte Carlo 
method of Vokrouhlick\'{y} et al. (2006a, b, c), modified to account 
for new developments in our understanding of the YORP effect, to try to 
refine the preliminary estimates of the family age obtained in this section.

\section{Chronology}
\label{sec: chron}

Monte Carlo methods to obtain estimates of the family age and ejection 
velocity parameters were pioneered by Milani and Farinella (1994)
and improved by Vokrouhlick\'{y} et al. (2006a, b, c) 
for the Eos and other asteroid groups.   They were recently modified
to account for the ``stochastic'' version of the YORP effect (Bottke et al. 
2015), and for changes in the past values of Solar luminosity 
(Vokrouhlick\'{y} et al. 2006a) for a study of dynamical groups in the 
Cybele region (Carruba et al. 2015).  We refer the reader to the latter 
paper for a more in depth description of the method.  Essentially, the 
semi-major axis distribution of various fictitious families is evolved under
the influence of the Yarkovsky, both diurnal and seasonal versions, 
and YORP effect (and occasionally other effects such as close encounters 
with massive asteroids (Carruba et al. 2014) or changes in past solar 
luminosity values (Carruba et al. 2015).  The newly obtained distributions 
of a $C$-target function are computed with the relationship:

\begin{figure}
  \centering
  \centering \includegraphics [width=0.45\textwidth]{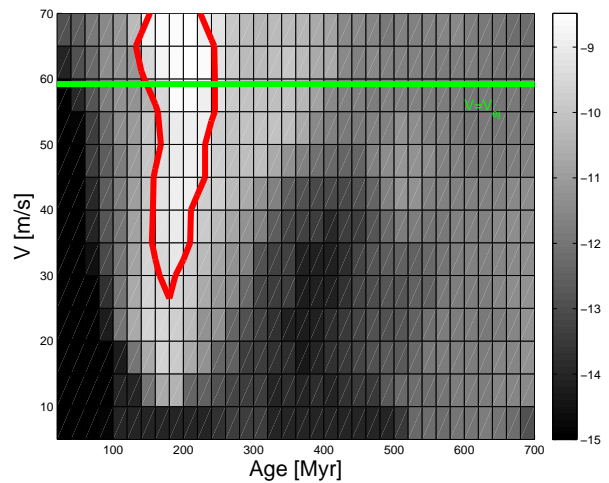}

\caption{Target function ${\psi}_{\Delta C}$ values in the ($Age,V_{EJ}$) plane 
for Erigone family. The horizontal green line displays the value of 
the estimated escape velocity from the parent body.}
\label{fig: cont_erigone}
\end{figure}

\begin{equation}
0.2H=log_{10}(\Delta a/C),
\label{eq: target_funct_C}
\end{equation}  

\noindent where $H$ is the asteroid absolute magnitude.
We applied this method to the Erigone group obtained in 
Sect.~\ref{sec: fam_ide}.  Fig.~\ref{fig: cont_erigone} 
displays ${\psi}_{\Delta C}$ values in the ($Age,V_{EJ}$) plane for this 
family.  Values of the $V_{EJ}$ parameter
have to be lower than the estimated escape velocity from the parent
body (Bottke et al. 2015), 59.2 m/s for the case of the Erigone family.  
Using ${\psi}_{\Delta C} = 9.4$ (red line in Fig.~\ref{fig: cont_erigone}), 
corresponding at a confidence level of 81.2\% 
to the two distribution being compatible (Press et al. 2001, there 
were 12 intervals in the $C$ distribution), the Erigone family should be 
$190^{+50}_{-40}$~My old, with $V_{EJ} = 55^{+5}_{-30}$~m/s.
As discussed in the introduction, Vokrouhlick\'{y} et al. (2006c) used 
a Yarko-Yorp approach to find an age of $280^{+30}_{-50}$~My.  The slightly 
smaller value found in this work could be caused by the fact that 
we are including the stochastic version of the YORP effect.  Our estimate
seems to be more in agreement with that of Bottke et al. (2015),  
$170^{+25}_{-30}$~My, that included this effect, and with the age estimate
of Spoto et al. (2015), who found an age of $220^{+60}_{-80}$ My with an 
independent method\footnote{In Spoto et al. (2015) the Erigone family is 
considered as the merger of the Erigone family and the 5026 Martes group.}.  
Overall, there seems to be a consensus in the literature 
that the Erigone family should be a relatively young C-type cratering family.

In the next section we will study the constraints that can be put 
on the Erigone family age by dynamics inside the $z_2$ secular resonance.

\section{Dynamics inside the $z_2$ secular resonance}
\label{sec: z2_dyn}

About 14.4\% of the members of the Erigone family identified in the 
$(a,e, sin(i))$ domain are currently in librating states of the $z_2$ 
secular resonance.  While well below 50\% of the family membership, there 
is still quite a significant population of Erigone members in librating 
states to allow to infer some constraints on the dynamical evolution and age of
the Erigone group.

\begin{figure*}

  \centering
  \begin{minipage}[c]{0.5\textwidth}
    \centering \includegraphics[width=3.5in]{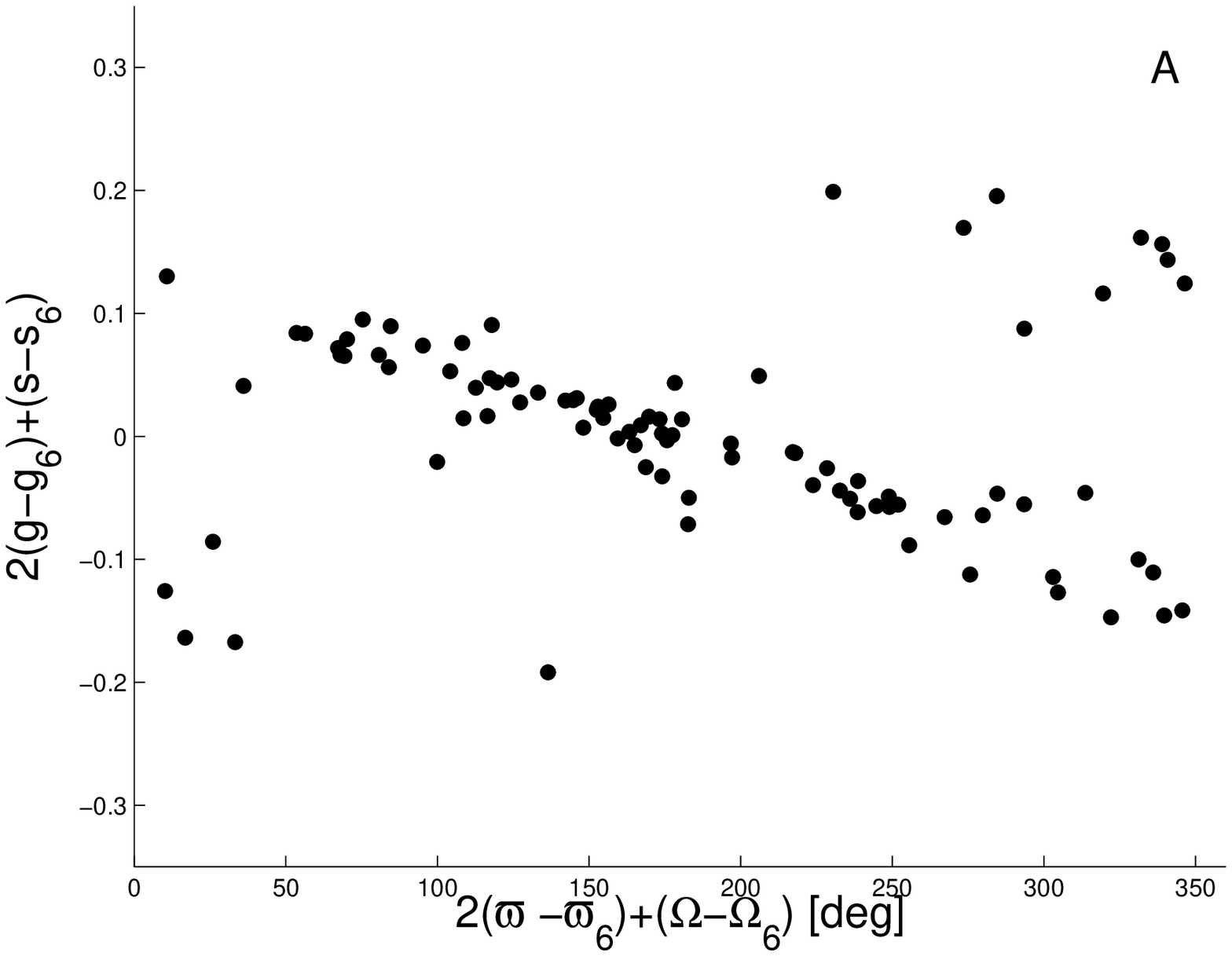}
  \end{minipage}%
  \begin{minipage}[c]{0.5\textwidth}
    \centering \includegraphics[width=3.5in]{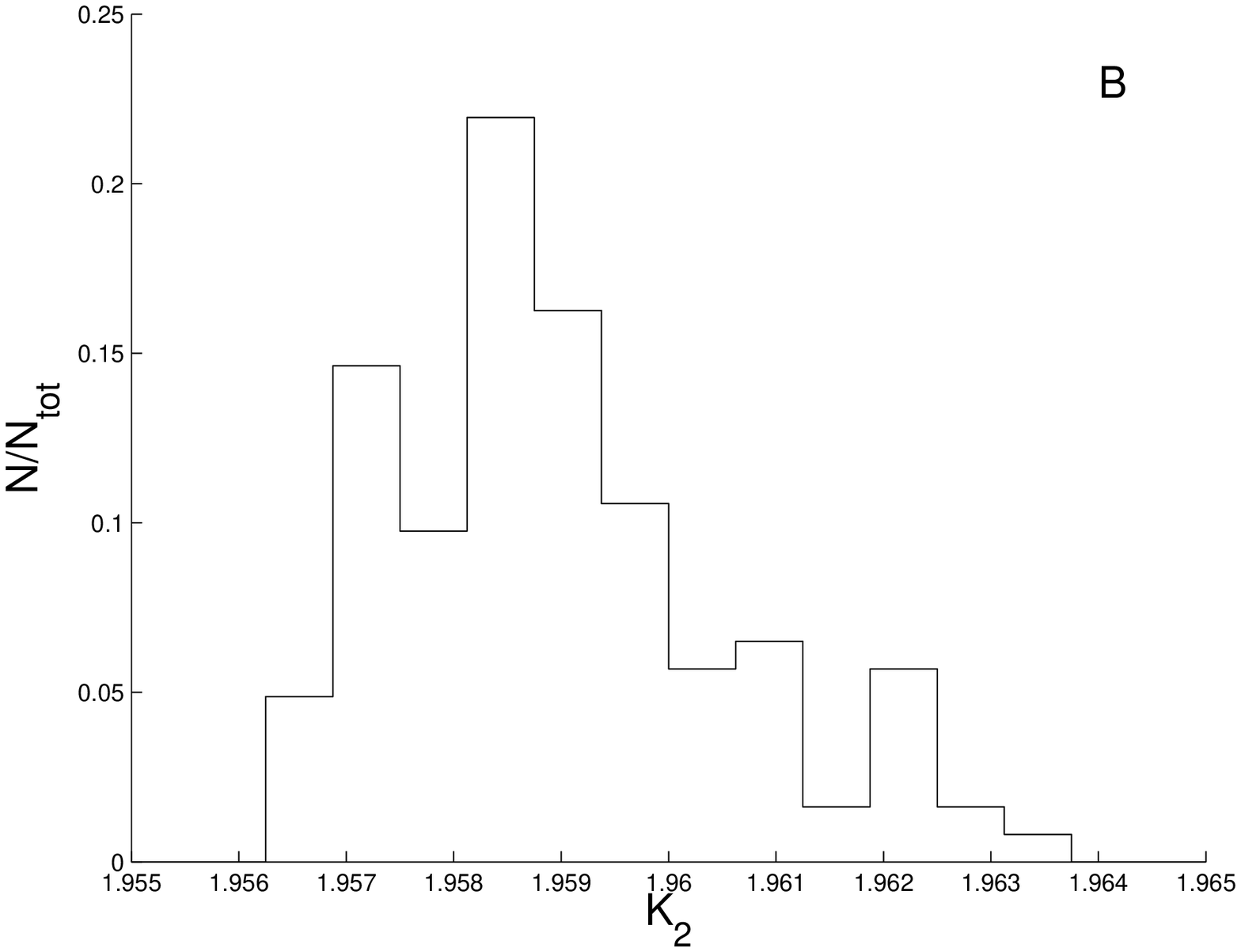}
  \end{minipage}

\caption{Panel A: $z_2$ librating members of the $(a,e,i)$ (full black 
dots) Erigone family, displayed in a plane where the x-axis is the 
$z_2$ resonant argument $2(\varpi -{\varpi}_6) +(\Omega-{\Omega}_6)$ and
the y-axis is the associated frequency $2(g-g_6)+(s-s_6)$. 
Panel B: normalized distribution of the quasi-integral
${K_2}^{'}=\sqrt{1-e^2}(3-\cos(i))$ values for the Erigone 
$(a,e,i)$ (black line) members currently inside the $z_2$ resonance.} 
\label{fig: K2_Z2}
\end{figure*}

First, we analyzed the current dispersion of librating asteroids
in the plane $(\sigma,\frac{d\sigma}{dt})$, where $\sigma$ is the $z_2$ 
resonant argument $2(\varpi -{\varpi}_6) +(\Omega-{\Omega}_6)$, and 
$\frac{d\sigma}{dt}$ is the associated frequency $2(g-g_6)+(s-s_6)$.  
The procedure followed to compute the quantities in Fig.~\ref{fig: K2_Z2}, 
panel A is as follows. The orbital elements results of the numerical 
simulation over 30 Myr are used to obtain equinoctial, 
non-singular element of the form
$(e\cdot \cos{\varpi}, e\cdot \sin{\varpi})$, and
$(\sin{(i/2)}\cos{\Omega}, \sin{(i/2)}\sin{\Omega})$.
The equinoctial elements of the test particles
and of Saturn are then Fourier filtered to obtain the $g, s, g_6, s_6$
frequencies and their associated phases.  The frequencies are then
plotted on the ordinates and their phases are used to
construct the resonant angle $\sigma$, for the 92 objects that 
remained in librating states for the whole lenght of the simulation.
  
\begin{figure*}
  
  \centering
  \begin{minipage}[c]{0.5\textwidth}
    \centering \includegraphics[width=2.5in]{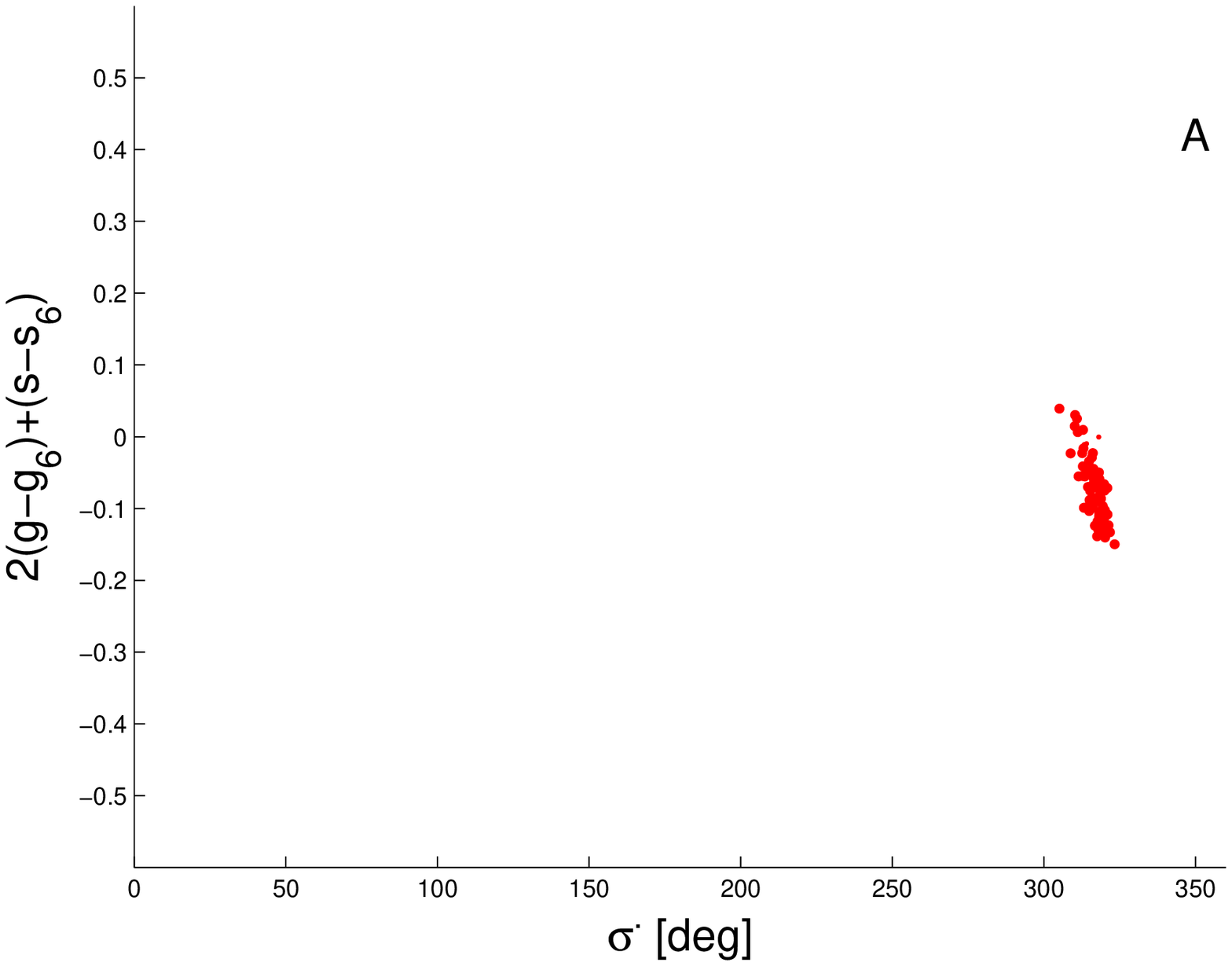}
  \end{minipage}%
  \begin{minipage}[c]{0.5\textwidth}
    \centering \includegraphics[width=2.5in]{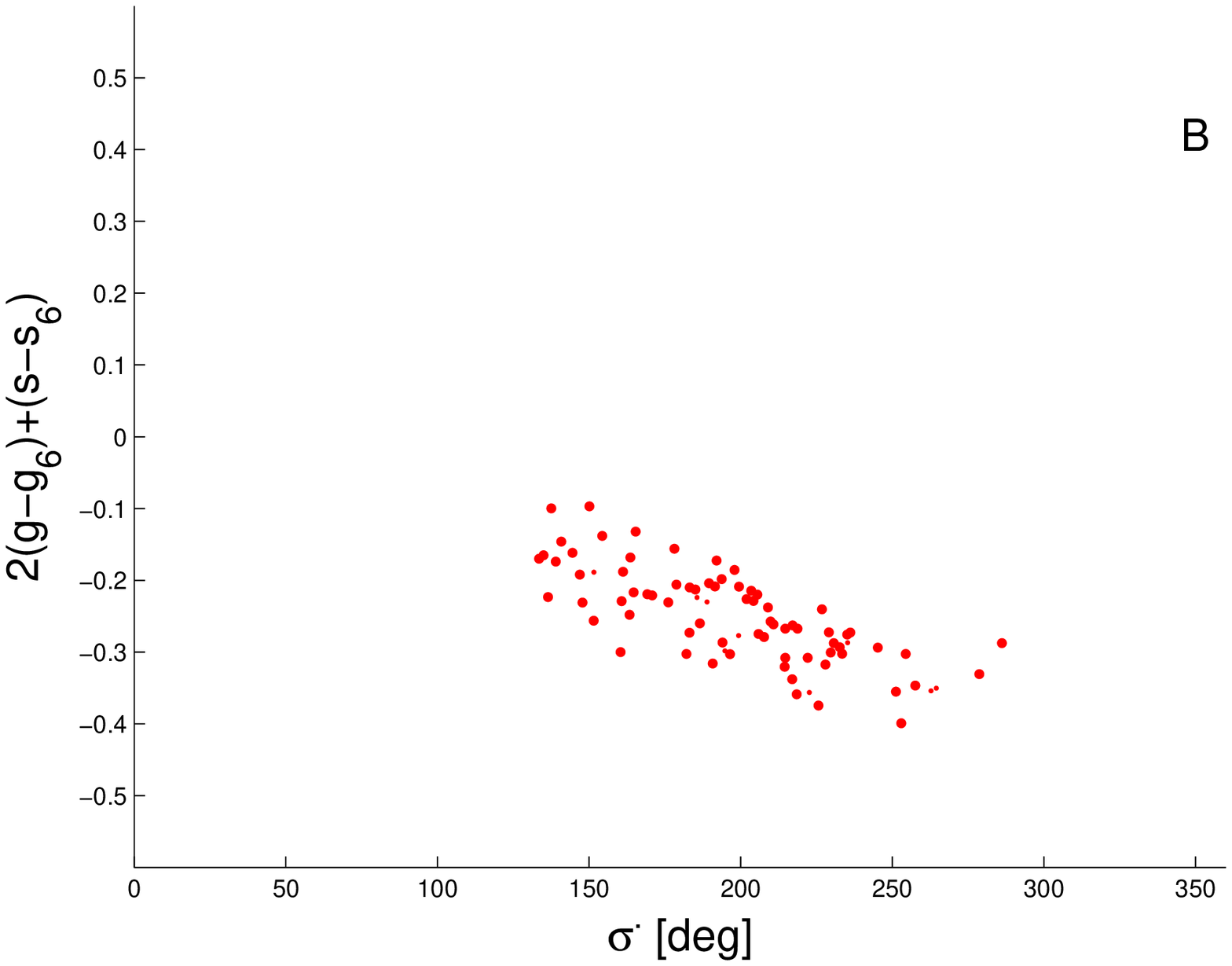}
  \end{minipage}

  \centering
  \begin{minipage}[c]{0.5\textwidth}
    \centering \includegraphics[width=2.5in]{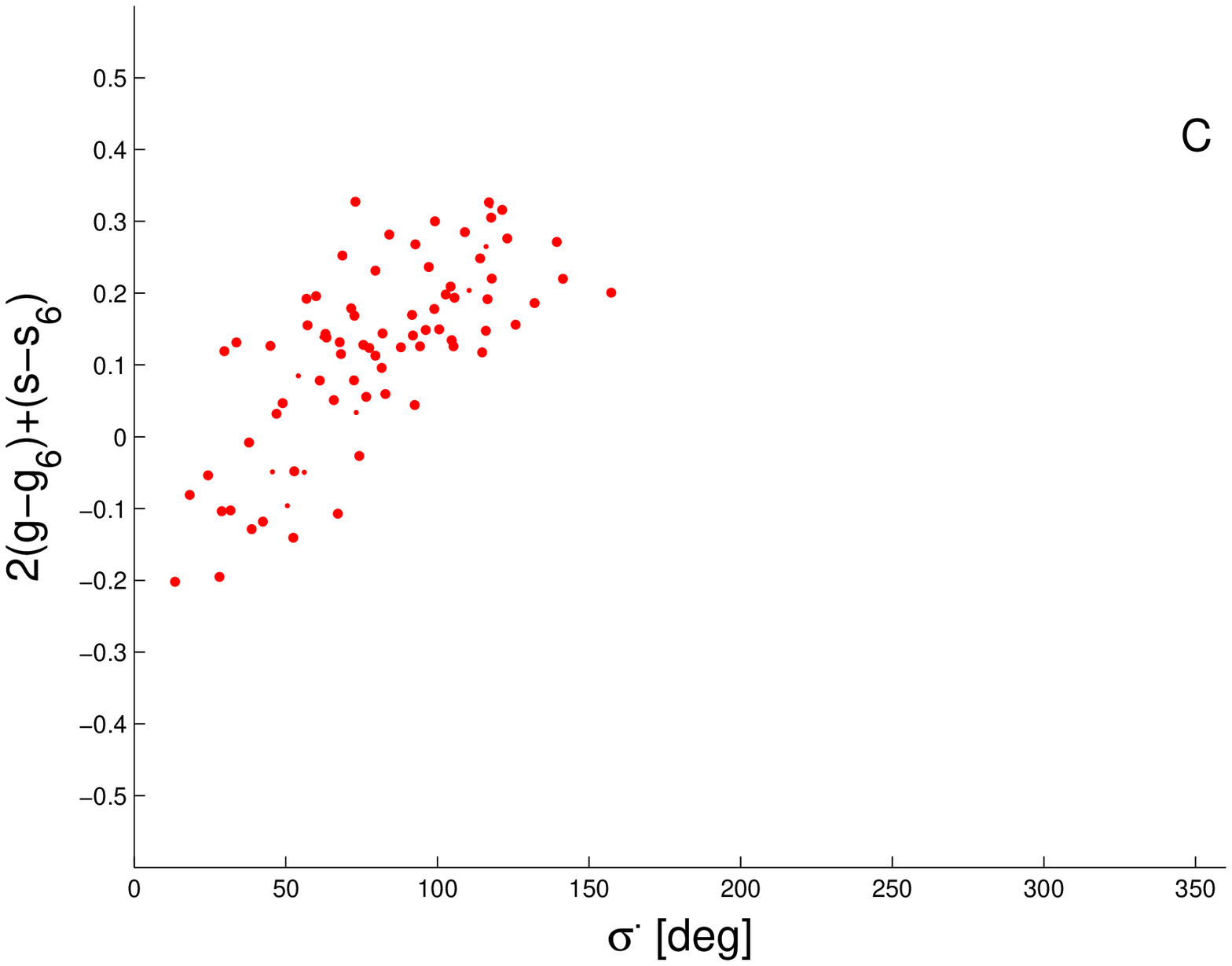}
  \end{minipage}%
  \begin{minipage}[c]{0.5\textwidth}
    \centering \includegraphics[width=2.5in]{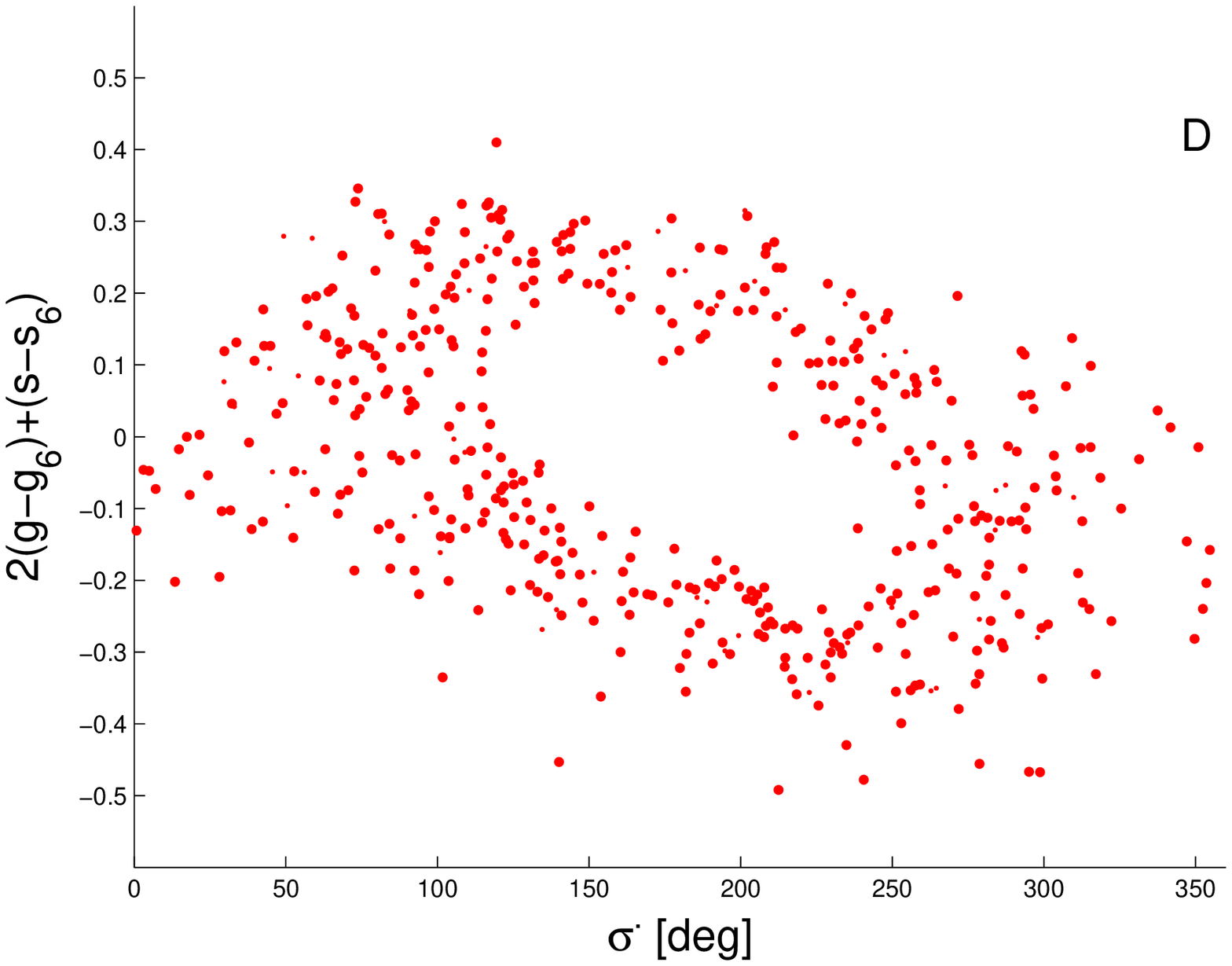}
  \end{minipage}

\caption{Snapshots of the numerically integrated evolution of clones
of Erigone family member 8089 Yukar (red dots) projected onto 
the $z_2$ resonance variables: the state after t = 1.2 My (panel A), 
after 3.7 My (panel B), after 6.1 My (panel C), and all data points 
between 3.7 and 9.8 My, panel D).}
\label{fig: z2_snap}
\end{figure*}

The reader may notice in Fig.~\ref{fig: K2_Z2}, panel A, that 
no resonant asteroid was found at values of $2(g-g_6)+(s-s_6)$ greater than 
0.3 arcsec/yr or lower then -0.3 arcsec/yr, so confirming the criteria
of Carruba (2009) used in Sect.~\ref{sec: sec_dyn}.  One can also notice 
that, as for the case of the ${\nu}_6$ (Carruba and Morbidelli 2011) and $z_1$
resonances (Vokrouhlick\'{y} et al. 2006b, Carruba 2009), there is an 
equilibrium point at $\sigma = 180^{\circ}$.  Contrary to the cases of 
the ${\nu}_6$ and $z_1$, data has a poor signal-to-noise ratio and is
difficult to interpret.  The resonant behavior of a single particle will
be better displayed later on in this section (see Fig.~\ref{fig: z2_snap}).

Fig.~\ref{fig: K2_Z2}, panel B, deals with conserved quantities of the $z_2$ 
secular resonance.  It is possible to show (see Carruba and Morbidelli 2011, 
Vokrouhlick\'{y} et al. 2006b, Carruba 2009, and calculations therein) that
${K_2}^{'} = \sqrt{1-e^2}(1-\cos{i})$ is conserved for the ${\nu}_6$
resonance, ${K_2}^{'} = \sqrt{1-e^2}(2-\cos{i})$ is preserved
for the $z_1$ resonance, and ${K_2}^{'} = \sqrt{1-e^2}(3-\cos{i})$ 
is conserved for the $z_2$ one.  Fig.~\ref{fig: K2_Z2}, panel B,
displays normalized distributions of the quasi-integral ${K_2}^{'}$
for the $(a,e,i)$ (black line), and $(n,g,2g+s)$ (green line) groups.
Since less than 50\% of the Erigone members are in $z_2$ librating states,
we cannot use the conservation of this quantity to set constraints
on the original ejection velocity field, as done for the Tina, Agnia
and Padua family.  Yet, checking its behavior as a function of time,
may show how well this quantity is actually preserved for the 
$z_2$ resonance, and if it could be possibly used for other $z_2$ 
resonant groups, should they ever be discovered.  

To further investigate dynamics inside the $z_2$ resonance we devised this 
numerical experiment:  we started with a synthetic Erigone family as a 
tightly compact cluster in $a,e,sin(i)$ and $\sigma$.  This family is initially 
a compact cloud as shown in the first panel of Fig.~\ref{fig: z2_snap}.
To better define the time it takes for the cloud to disperse, the 
fake family is composed by  81 ``clones'' produced by adding and subtracting 
small values to the eccentricity (0.0001) and inclination ($0.001^{\circ}$) of 
8089 Yukar, the lowest numbered asteroid in our sample in a $z_2$ librating 
state. 

\begin{figure*}

  \centering
  \begin{minipage}[c]{0.5\textwidth}
    \centering \includegraphics[width=2.5in]{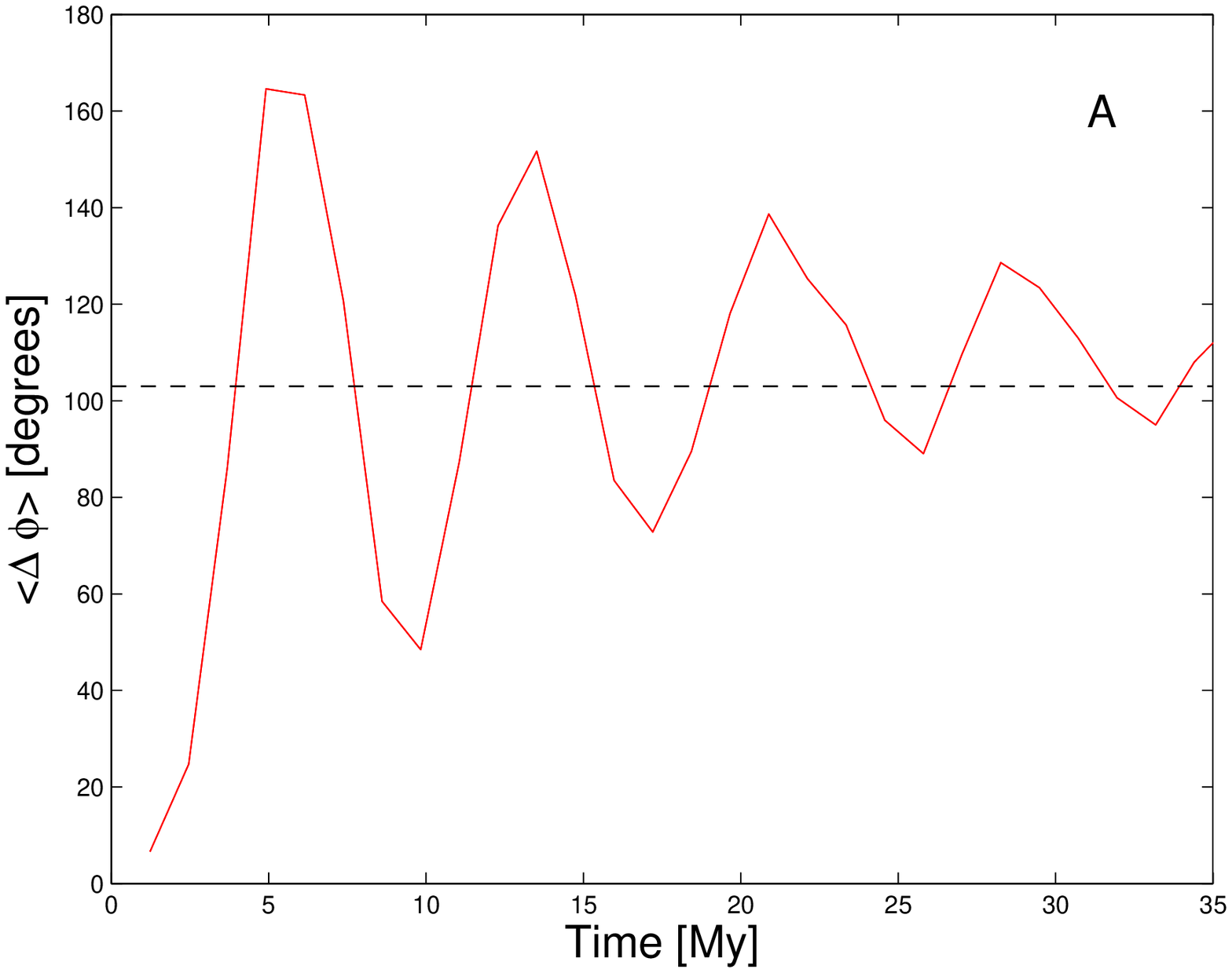}
  \end{minipage}%
  \begin{minipage}[c]{0.5\textwidth}
    \centering \includegraphics[width=2.5in]{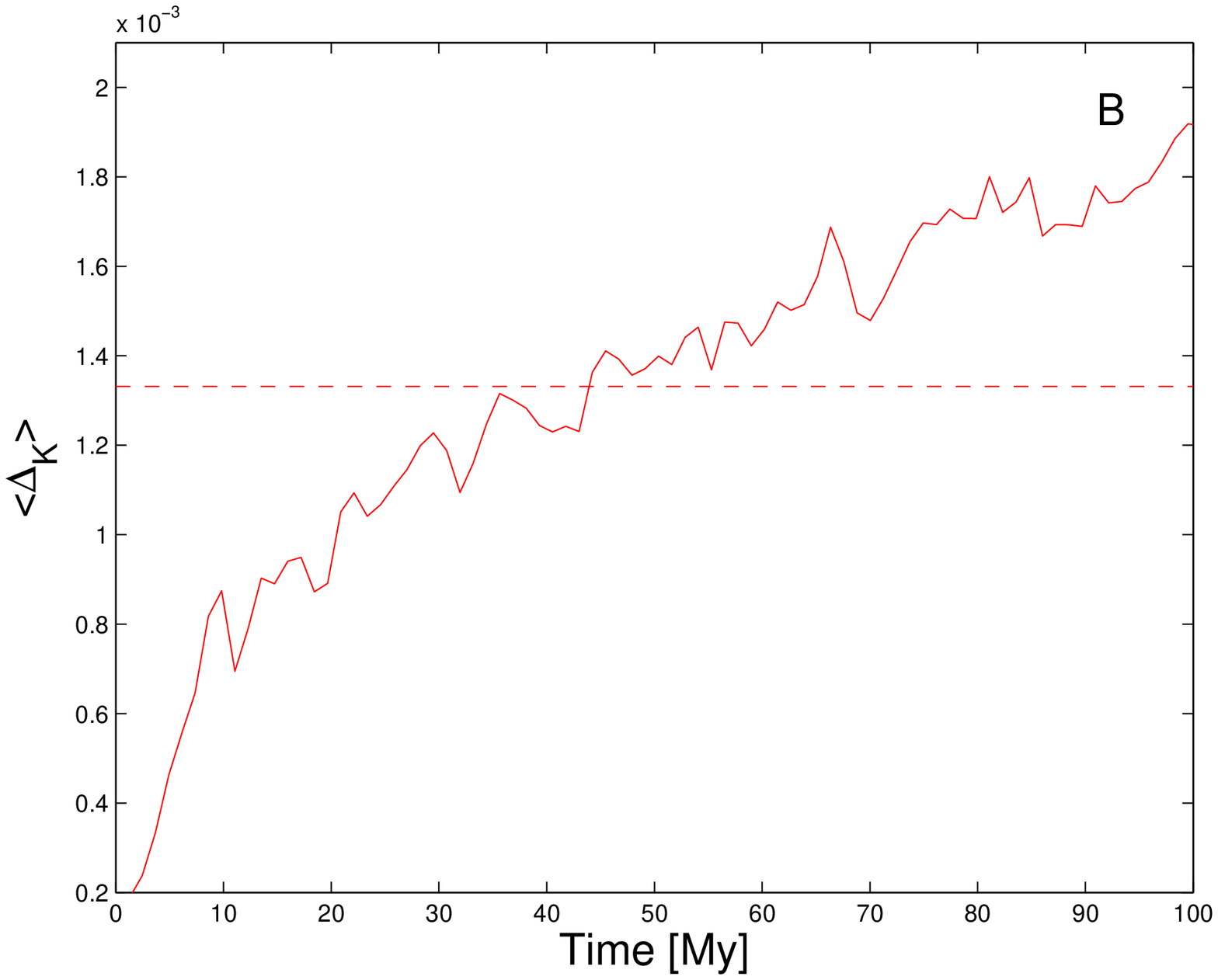}
  \end{minipage}

    \caption{Panel A: temporal evolution of $D^2_{\Phi}$ of 
Eq.~\ref{eq: D_phi} for the synthetic Erigone family.  Panel B: temporal
evolution of $D_K$, dispersion of the $K_2^{'}$ variable.  The dotted line 
displays the mean value of $D_K$ during the simulation.}
\label{fig: Dphi}
\end{figure*}

Fig.~\ref{fig: z2_snap} shows the evolution of the synthetic family as 
tracked in the space of the $z_2$ resonance variables $(\sigma,d\sigma/dt)$.
The last panel show all data points between 3.7 and 9.8 My, a period that
covers an entire $z_2$ libration cycle.  
After $\simeq$ 12.3 My the initially tight cluster becomes uniformly 
dispersed along the separatrix of the $z_2$ resonance.  Notice that,
contrary to the cases of the ${\nu}_6$ and $z_1$ secular resonances,
while the period of libration remains approximately constant, 
the amplitude changes in a chaotic fashion.
After a few librations cycles particles with very close initial conditions may
find themselves quite apart in the $(\sigma,d\sigma/dt)$ plane, as 
indeed observed in Fig.~\ref{fig: K2_Z2}, panel A.   
To quantitatively describe the distribution of bodies along the $z_2$ 
separatrix, we used the polar angle $\Phi$ in the $(\sigma,d\sigma/dt)$ 
plane described in  Vokhrouhlick\'{y} et al. (2006b)\footnote{We used
a scaling that maps a $(0^{\circ},360^{\circ})$ interval of
$\sigma$ and (-1,1) ``/yr interval in $d\sigma/dt = g+s-g_6 - s_6$ into 
common intervals (-1,1).}. At each step of the numerical simulation, 
we computed the dispersion $D²_{\Phi}$ in the polar angle $\Phi$ defined
as:

\begin{equation}
D^2_{\Phi}=\frac{1}{N(N-1)}{\displaystyle\sum}_{i \neq j}({\Phi}_i-{\Phi}_j)^2,
\label{eq: D_phi}
\end{equation}

\noindent
where N = 81 is the number of integrated bodies and ${\Phi}_i$ is the
polar angle of the $i$-th body (i = 1,...,N).  Since we started with 
a compact cluster, $D^2_{\Phi}$ is initially small ($\simeq 6.61^{\circ}$),
but grows with time because of the differential libration of the bodies
in the resonance (Fig.~\ref{fig: Dphi}, panel A).   After only 
$\simeq$ 12 My, i.e., about two libration cycles of the $z_2$ resonance
for (8089) Yukar, the value of $D^2_{\Phi}$ saturates at 
$\simeq 103^{\circ}$, which corresponds to an uniform distribution of 
bodies along a circle (Vokrouhlick\'{y} et al. 2006b).  This result 
suggests that any information about initial distributions of orbits
inside the $z_2$ resonance is lost very quickly, on a timescale of the 
order of 10 Myr.

Fig.~\ref{fig: Dphi}, panel B, displays the time evolution of the 
dispersion $D_K$ of the $K_2^{'}$ quantity.  Contrary to the case
of the $z_1$ resonance, where $D_K$ was nearly constant over 
the entire integration time span of 100 My and perturbations occurred 
only when some of the bodies left the $z_1$ resonance, 
here perturbations of the $K_2^{'}$ are quite significant.  The percentual
change in $D_K$, defined as the standard deviation of $D_K$ 
divided by its averaged value, was 30.1\%.   Results are similar when
non-gravitational forces such as the Yarkovsky and YORP effects are
considered, and will not be here discussed for the sake of brevity.
This result shows that it is not easily feasable to use 
the observed $K_2^{'}$ distribution to infer 
quantitative information about the initial velocity field, as done 
for instance for families inside the $z_1$ resonance, such as the 
Agnia and Padua groups.  This is not a big issue for the Erigone 
family, since the majority of its members are not currently in 
$z_2$ librating states, but is a fact to be considered if a dynamical
group with a majority of its members in such configuration will be 
discovered in the future. 

The next section will deal with the dynamical evolution of the Erigone
family as whole, and on what constraints can be set by studying its
interaction with the $z_2$ secular resonance. 

\section{Dynamical evolution of the Erigone family}
\label{sec: dyn_evol}

To study the dynamical evolution of the Erigone family members, 
we performed simulations with 
the $SYSYCE$ integrator (Swift$+$Yarkovsky$+$Stochastic
YORP$+$Close encounters) of Carruba et al. (2015), modified to also
account for past changes in the values of the solar luminosity.
The numerical set-up of our simulations was similar to the one discussed
in Carruba et al. (2015): we used the optimal values of the Yarkovsky 
parameters discussed in Bro\v{z} et al. (2013) for C-type asteroids,
the initial spin obliquity was random, and normal reorientation timescales 
due to possible collisions as described in Bro\v{z} (1999) were 
considered for all runs. We integrated our test particles  
under the influence of all planets, and obtained synthetic proper elements
with the approach described in Carruba (2010).

\begin{figure}
  \centering
  \centering \includegraphics [width=0.45\textwidth]{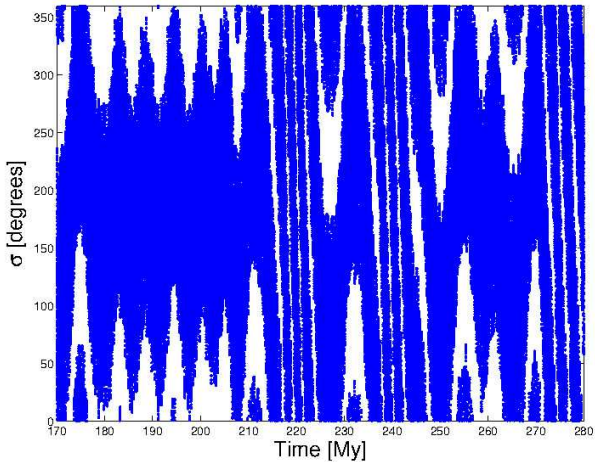}

\caption{Time behavior of the $z_2$ resonant argument of a simulated
particle between 170 and 280 My.}
\label{fig: res_ang_long}
\end{figure}

We generated a fictitious family of 854 objects 
with the ejection parameter $V_{EJ}$ found in Sect.~\ref{sec: chron}, and 
integrated this group over 400 My, 
well beyond the maximum possible value of the age of the Erigone family.  
We then analyzed the resonant argument $\sigma$ of the $z_2$ secular
resonance for all simulated particles.  The behavior of this angle could 
be quite complex: because of the drift in semi-major axis caused by 
non-gravitational forces, asteroids can be temporarily captured in the
$z_2$ resonance, then escape to circulating orbits, then, should the
direction of the drift change, be captured again, etc.  
Fig.~\ref{fig: res_ang_long} displays the time dependence of 
the $z_2$ resonant argument of a simulated particle that alternated
between phases of libration and circulation in the $z_2$ secular
resonance.  Overall, while the number of $z_2$ librators 
should change with time, if the population of librators 
from the Erigone family is in a steady-state, this number
should fluctuate around the median value, with fluctuation of the
order of one standard deviation.  Similar behavior is for instance
observed for the population of asteroids currently inside the
M2:1A mean-motion resonance (Gallardo et al. 2011).  The minimum
time needed to reach a steady-state could therefore be used
to set constraints on the age of the Erigone family.

\begin{figure}
  \centering
  \centering \includegraphics [width=0.45\textwidth]{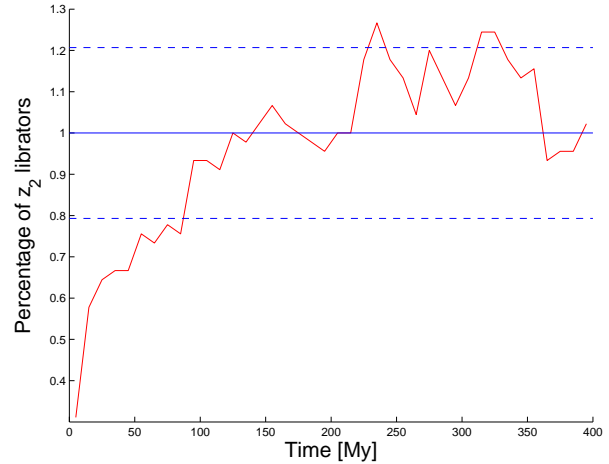}

\caption{Fraction of simulated Erigone family members in $z_2$ librating 
states as a function of time, normalized with respect to the median 
value.  The horizontal blue line displays the 
median percentage of objects in the $z_2$ states, dashed lines display
levels of median fraction plus or minus its standard deviation.}
\label{fig: z2_dyn_constr}
\end{figure}

We analyzed the resonant angle of all simulated particles, and computed
the fraction of family members in $z_2$ resonant states as a function
of time.  Fig.~\ref{fig: z2_dyn_constr} displays our results:  the
number of $z_2$ librators fluctuates with time, but reaches its median
value after 125 My (after $\simeq$ 90 My if we consider
the median value plus or minus the standard deviations as an estimate
of the error, blue dashed lines in Fig.~\ref{fig: z2_dyn_constr}).  
We believe that this result is 
of quite significance, since i) sets a lower limit on the age of the 
Erigone family, that is compatible with this family age estimates obtained 
in this and other works, and, ii) most importantly, is obtained 
using a new technique, purely based on secular dynamics, not previously 
used in the literature.

\section{Conclusions}
\label{sec: conc}

In this work we:

\begin{itemize}

\item Computed proper elements for a dynamical map and real asteroids
in the Erigone family region.  The Erigone family is characterized by its
interaction with the $z_2$ secular resonance, and we found synthetic
proper elements for 4717 asteroids.

\item Studied the local secular dynamics.  There are 367 asteroids
currently in $z_2$ librating states.  Other important non-linear secular
resonances in the area are the $3{\nu}_6-{\nu}_5$, a $g-$type resonance,
with 75 librators, and the ${\nu}_5+2{\nu}_{16}$, a $(g+2s)-$type resonance,
with 38 librators.

\item Revised current knowledge on the taxonomy and physical properties 
of local asteroids.  The Erigone region is dominated by the C-type 
Erigone family, but with significant mixing of other S and V type objects.
Except for the family-less S-type 60 Echo, the only other object with a 
mass larger than $10^{17}~kg$ in the region is 163 Erigone itself, with 
other bodies being most likely either members of the Erigone or of 
other families.

\item Identified the Erigone family in the domain of proper $(a,e,sin(i))$
elements and proper $(n,g,g+s)$ and $(n,g,2g+s)$ frequencies, the latter
domain being the most efficient in identifying objects in $z_2$ librating
states.  After eliminating taxonomical and dynamical interlopers, we 
obtained an $(a,e,sin(i))$ family of 854 members, 14.4\% of which in 
$z_2$ librating states.  

\item Obtained estimates of the Erigone family age with a Yarko-Yorp 
Monte Carlo method that also accounts for the ``stochastic'' version of 
the YORP effect, and for changes in the past values of Solar luminosity 
(Carruba et al. 2015).  At a 81.2\% confidence level, the Erigone family 
should be $190^{+50}_{-40}$~My old, with $V_{EJ} = 55^{+5}_{-30}$~m/s.
Results are in agreement with those of other groups (Vokrouhlick\'{y} et 
al. 2006c, Bottke et al. 2015, Spoto et al. 2015).

\item Studied the dynamics inside the $z_2$ secular resonance.  We observe 
one stable equilibrium point at $\sigma = 180^{\circ}$ in 
the $z_2$ resonance plane $(\sigma, \frac{d\sigma}{dt})$, with 
$\sigma = 2(\varpi -{\varpi}_6) +(\Omega-{\Omega}_6)$ the resonant angle
of the $z_2$ resonance.  Current members of the $(a,e,sin(i))$ and 
$(n,g,2g+s)$ Erigone families are fully dispersed around the  
equilibrium point.  Since the minimum time to achieve such configuration
is $\simeq 12$ My, this sets a lower limit on the Erigone family age.

\item Studied the dynamics of the Erigone family as a whole, and 
showed that the minimum time to inject into $z_2$ librating states
a steady-state population is 90 My, which sets another constraint on 
the family age.  

\end{itemize}

Overall, our estimate of the Erigone family age is compatible with
values in the literature.  More important, by studying the very 
interesting dynamics inside the $z_2$ secular resonance, we were able
to set two additional independent lower limits to the family age.
These new tools based on secular dynamics of a resonant family constitute,
in our opinion, one of the main results of this work.

\section*{Acknowledgments}
We are grateful to an anonymous reviewer for comments and suggestion
that improved the quality of this paper.  This paper was written 
while the first author was a visiting scientist
at the SouthWest Research Institute (SWRI) in Boulder, CO, USA. 
We would like to thank the S\~{a}o Paulo State Science Foundation 
(FAPESP) that supported this work via the grants 14/24071-7, 
14/06762-2, 2013/15357-1, and 2011/08171-3, and the Brazilian 
National Research Council (CNPq, grants 312313/2014-4 and 312813/2013-9).
This publication makes use of data products from the Wide-field 
Infrared Survey Explorer, which is a joint project of the University 
of California, Los Angeles, and the Jet Propulsion Laboratory/California 
Institute of Technology, funded by the National Aeronautics and Space 
Administration.  This publication also makes use of data products 
from NEOWISE, which is a project of the Jet Propulsion 
Laboratory/California Institute of Technology, funded by the Planetary 
Science Division of the National Aeronautics and Space Administration.

\bsp

\label{lastpage}

\end{document}